
\magnification=1200
\tolerance=10000
\newcount\eqnumber
\eqnumber=1

\def\eqnam#1{\xdef#1{\the\eqnumber}}
\def\new{\the\eqnumber\global\advance\eqnumber by 1}
\footline={\ifnum\pageno>1\hfil\folio\hfil\else\hfil\fi}
\pageno=1
\rightline{IASSNS-HEP-92/42}
\rightline{BNL-47671}
\rightline{June 1992}
\vskip.5in
\centerline{\bf Low temperature expansion for the 3-d Ising Model}
\bigskip
\centerline{\it by}
\bigskip
\bigskip
\centerline{\bf Gyan Bhanot}
\medskip
\centerline{\sl School of Natural Sciences}
\centerline{\sl Institute for Advanced Study}
\centerline{\sl Princeton, NJ 08540}
\medskip
\centerline{\it and}
\medskip
\centerline{\sl Thinking Machines Corporation}
\centerline{\sl 245 First Street, Cambridge MA 02142}
\medskip
\medskip
\centerline{\bf Michael Creutz}
\medskip
\centerline{\sl Physics Department}
\centerline{\sl Brookhaven National Laboratory}
\centerline{\sl Upton NY 11973}
\medskip
\medskip
\centerline{\bf Jan Lacki}
\medskip
\centerline{\sl School of Natural Sciences}
\centerline{\sl Institute for Advanced Study}
\centerline{\sl Princeton, NJ 08540}
\bigskip

\centerline{ABSTRACT}
\medskip
\baselineskip=18pt
{\narrower We compute the weak
coupling expansion for the energy of the three dimensional Ising
model through 48 excited bonds. We also compute the magnetization
through 40 excited bonds. This was achieved via a recursive
enumeration of states of fixed energy on a set of finite lattices.
We use a linear combination of lattices with a generalization of
helical boundary conditions to eliminate finite volume effects.

}

\vfill\eject
Expansions about either infinite or vanishing coupling are a major
technique for the study of critical properties of statistical systems
and field theories. These series usually involve a diagrammatic
analysis which becomes rapidly more complex as the order increases.
Thus it would be interesting to have an automated technique for the
generation of the relevant terms.

Here we consider generating the low temperature or weak coupling
expansion for discrete systems.  Our approach does not involve
explicit graphs, but relies on a recursive computer enumeration of
configurations.  We illustrate the approach on the three dimensional
Ising model.

The method uses a procedure of Binder [1] for the  explicit solution
of discrete models on small lattices.  In Ref.~[2] these ideas were
further developed. Ref.~[3] explored extracting the low temperature
series.  This paper adds further tricks to obtain additional terms.
This extends the  low temperature series to an order comparable to
existing high temperature expansions [4].

 \def\refmc{3}

We consider the Ising model on a three dimensional simple cubic
lattice.  On each site $i$ is a spin $\sigma_i$ taking the values
$\pm 1$.  The system energy is
 $$
 E=\sum_{\{i,j\}}(1-\sigma_i \sigma_j)-H\sum_i \sigma_i \eqno(\new)
 $$
where the first sum is over all nearest neighbor pairs of spins, each
pair being counted once.  Temporarily we set the applied field $H$ to
zero. The partition function is the sum of the Boltzmann weight over
all configurations
 $$
Z=\sum_{\{\sigma\}} e^{-\beta E} \eqno(\new)
$$
Sorting configurations by energy, we rewrite this as a sum over $E$.
We define $P(E)$ to be the number of states with a given energy $E$.
Thus, we have
 $$
 Z=\sum_{E=0}^{6N}P(E) u^{E/2}\eqno(\new)
 $$
where $N$ is the number of sites and $u=e^{-2\beta}$.

We compute the coefficients $P(E)$ exactly on small systems.   We
recursively assemble the system one site at a time.  The method
enables us to build up a lattice with arbitrary length in one of the
three dimensions.  Intermediate stages require an explicit
enumeration of all exposed two dimensional slices.  This effectively
reduces the computational complexity to that of a system of one less
dimension.

The starting point is a list of all states and corresponding energies
for a single transverse layer of the lattice.  All spins outside this
layer are frozen to the same value; that is, the boundary conditions
in the longitudinal direction are cold.  Spins are then sequentially
freed to build up the lattice in this third direction.  We
store the exact number of states of
any given energy and specified exposed top layer.  Storing the top
layer in the bits of an integer $I$, we define $p(E,I)$ to be this
count.  When a new spin or set of spins is added, we obtain the new
counts $p^\prime(E,I)$ as a sum over the old counts
 $$
 p^\prime(E,I)=\sum_{I^\prime} p(E-\Delta(I,I^\prime),I^\prime).
\eqno(\new)
 $$
Here $I^\prime$ can differ from $I$ only in the
bits representing the newly covered spins, and $\Delta(I,I^\prime)$
is the change in energy from any newly changed bonds.   For the
present analysis we add the spins one at a time.  Thus, the sum in
the above equation is only over two terms, representing the two
possible values for the newly covered spin. After the lattice is
grown, a sum over the top layers gives the resulting $P(E)=\sum_I
p(E,I)$  We always continue the recursion sufficiently to avoid
finite size errors in the longitudinal direction.

As the temperature goes to zero, so does the variable $u$.
Thus Eq.~(3) is the low temperature expansion for $Z$.  From
it, we compute the series for the average energy per
site,$\langle E \rangle =2\left(u{\partial \over \partial u}\right)
\log(Z).$
Comparing this expectation before and after adding the last spin,
we obtain the average energy per new site.  Expanding
in powers of $u$ gives
 $$
\langle E/N \rangle = \sum_j e_j u^j\eqno(\new)
 $$
We are interested in the coefficients $e_j$ in the infinite volume
limit.  One of our primary results, given in Table I, is the values
of these coefficients through $j=48$.

At zero temperature $(\beta=\infty)$ the only surviving states
have all spins parallel.  As the temperature increases, groups of
spins can flip in this uniform background.  Enumerating the possible
combinations gives a diagrammatic method to obtain the low
temperature coefficients [5]. Note that any enclosed group of flipped
spins always involves an even number of excited bonds.  Thus the
expansion only contains even powers of $u$.  We use a combination of
periodic transverse and cold longitudinal boundary conditions to
ensure this remains true on our finite systems.

On a periodic lattice of size $n\times n\times n$, the order
to which the weak coupling expansion for $\langle E/N \rangle$ will
agree with the infinite volume limit is $4n-2$.  A line of $n$
flipped spins can wrap around the lattice and  have energy $4n$
rather than the $4n+2$ it would have in infinite space.  This order
can be increased via boundary conditions to require more spins to be
flipped to wrap around the lattice.  Ref.~[\refmc] showed a version
of helical boundaries whereby an $n$ by $n$ transverse slice is
mimicked with only $[(n^2+1)/2]$ sites.  Here we extend this idea to
include the helicity into the longitudinal direction.

We build our lattices one site at a time; so, it is natural to imagine
the sites lying in a line.  We do not, however, consider sequential
sites as nearest neighbors.  Instead, we introduce three integer
parameters $\{h_x,h_y,h_z)$ representing the distance along the line
to the nearest neighbor in the corresponding $x$, $y$, or $z$
direction.  Labeling sites in the sequence by their ordinal number $i$,
the nearest neighbors of site $i$ are at $i\pm h_x$, $i\pm h_y$ and
$i\pm h_z$.  For convenience, assume $h_x<h_y<h_z.$ With this
convention, all sites more than $h_z$ steps back in the chain are
covered.  Thus the recursion only requires us to keep explicit track
of the $h_z$ ``exposed" spins at the end of our chain.

A minimal closed loop on such a lattice consists of a number of
steps such that $ n_x h_x+n_y h_y+ n_z h_z=0,$
where $n_i$ represents the number of steps in the $i$th direction.
The length of such a loop is
 $n=\vert n_x\vert+\vert n_y\vert+\vert n_z\vert$.
On an infinite cubic lattice the only solution  to these equations is
the trivial case $n_i=0$.  On a finite lattice, any other solution
represents a finite size correction.  Flipping a chain of spins along
such a closed path generates a state with $4n$ excited bonds, and
creates a potential error in the series at that order.    For a simple
example, $(h_x,h_y,h_z)=(19,21,24)$  with
$(n_x,n_y,n_z)=(3,-5,2)$ gives a minimal loop of length 10 and will
give the series to the same order as a $10^3$ lattice.

Given parameters $(h_x,h_y,h_z)$, it is straightforward to
enumerate the minimal closed paths.  A different set of $h_i$
corresponds to a different set of such paths.  However, the
contribution to the coefficients $e_i$ from a particular path
is, by symmetry, independent of any permutations or sign changes in
the numbers $(n_x,n_y,n_z)$.  This allows us to
combine results from various size lattices to cancel
the contributions from particular closed loops.   For example,
consider loops of length 9.  The (16,18,21) lattice has a minimal
such loop with steps ${\bf n}=(3,2,-4)$, the $(16,17,21)$ lattice has
closed loops with steps $(1,4,-4)$ and $(5,-1,-3)$, the $(13,18,20)$
lattice has a closed loops with $(2,3,-4)$ and $(4,-4,1)$, and
finally the $(14,17,19)$ system has the loops $(3,2,-4)$ and
$(5,-3,-1)$.  If we combine the coefficients $e_i$ as obtained from
these lattices with weights $(2,1,-1,-1)$ respectively, then all
errors from the loops of length 9 cancel out.  This gives the series
to the same order as a lattice with the smallest loop having length
10, which otherwise requires at least 24 sites.

This procedure extends to cancel further loops.  For our calculation,
we assembled two lists of 10 lattices and the relative weights for
combining them to cancel all loops of length less than 13.  We ran
both combinations as a check on the error cancellations.  The first
set involved lattices with $h_z$ up to 23, and took about a day on an
IBM RS6000 workstation.  The second set, involving up to $h_z=24$,
used about half a day on a 32K Connection Machine and about a day of
Cray-YMP time.

Each lattice used had a minimal loop of length at least 9.
While a loop of length 13 has 52 excited bonds, we have a potential
error at order 50 because of the possiblity of a more
complex loop wrapping around the lattice simultaneously in a length 9
and a length 10 direction.  The minimal energy of such a possibility
is 50 excited bonds.  This is the limit on the order of the series
presented here.

During the recursive construction, each new count is a trivial sum of
just two terms, representing the two possibilities for the covered
spin. On the other hand, we must store counts for all energies up to
the maximum order desired as well as for all possible values of the
top $h_z$ spins of our helical lattice.  Thus, the primary
computational problem is storage.  To substantially reduce these
demands, we  performed the calculations modulo small integers so that
at intermediate stages the counts could be stored in one byte each.
This gives the final coefficients modulo the given integers.  After
multiple passes using mutually prime values for these modulos, we use
the Chinese remainder theorem to reconstruct the final series.

{}From our results we constructed the series for the ratio
 $$
 r_E ={(u{\partial \over \partial u}) E \over
  (u{\partial \over \partial u})^2 E} \eqno(\new)
 $$
As the first three $e_i$ vanish, this ratio is determined through
order $u^{42}$. $r_E$  should have a zero at the critical
point, with the slope at this zero equal to $2/\alpha$ where
$\alpha$ is the specific heat exponent.

The ratio test showed that the first singularity for the $E$ series
is unphysical and occurs near $u^2 = -1/3$. We therefore made a
conformal transform to new variables defined by $z=3u^2/(1+3u^2)$ to
map the interval $u^2=[-1/3,0]$ to $z=[-\infty,0]$ and the physical
interval $u^2=[0,\infty]$ to $z=[0,1]$. We then did a Pade analysis
in the variable $z$. The results of these are shown in Fig.~1 where
we plot a few stable Pade series for $r_E$ in the vicinity of the
expected singularity in $\beta$. There is a clear zero near
$\beta=0.22$, in good agreement with the Monte Carlo studies [6]
which give $\beta_c=0.22165....$ The average slope of the various
curves in Fig.~1 gives $\alpha=0.22$, which is about twice the
accepted value for this exponent. The small value for this quantity
makes its accurate determination difficult.

Extending these results to include the magnetic term in Eq. (1), we
augmented the counting to keep track of the number of flipped spins
as well as excited bonds.  This increases memory demands, so we
reduced the highest energy to 20 excited bonds, and worked on a
combination of smaller lattices with $h_z$ up to 19 to cancel closed
loops of length 8 through 10.  Assuming a spin up background, we
write
 $$
 {1\over 2} \langle 1-\sigma \rangle=\sum_{i,j} c_{ij} u^{2i}
\lambda^j
 \eqno(\new)
 $$
where $\lambda=\exp(-2\beta H)$. The coefficients $c_{ij}$ through 20
excited bonds are given in Table II.

Summing the numbers in Table II over rows gives the expansion in
$u^2$ for the magnetization at zero applied field.  In Fig.~2 we show
several Pade approximants for the ratio
 $$
r_\sigma= {{\langle \sigma \rangle} \over {
u {{\partial \langle \sigma \rangle} \over {\partial u }}}} \eqno(\new)
 $$
in the vicinity of the critical point.  Before making these
approximants, we made the same change of variables as used for
Fig.~(1).  These give an estimate for $\beta_c= 0.222$ and the
exponent $\hat\beta=0.31$, where $\hat\beta$ is defined by $\langle \sigma
\rangle\propto (\beta-\beta_c)^{\hat\beta}$ in the critical region.
These numbers are in reasonable agreement with the accepted values.

The method presented here should easily generalize to other discrete
systems.  The helical lattices used, as well as the combinations to
cancel out finite size errors, are independent of the Ising nature of
the spins. It is straightforward to introduce additional couplings,
although this will increase memory needs.  Some interesting
possibilities for futher exploration are gauge, Potts, and coupled
gauge-spin models in various dimensions.  Changing boundary
conditions should enable the study of interface properties.  A direct
application of these counting methods to the high-temperature or
strong-coupling limit may also be quite useful.  In Ref.~[7] similar
recursive methods were suggested as a means to study many fermion
systems.  A particularly challenging problem is the extension of
these ideas to theories with continuous spins.

\vskip .5in
{\bf\noindent Acknowledgements}
\vskip .2in

We thank Joseph Straley for discussions on turning finite lattice
partition functions into low temperature series. We also thank David
Atwood for discussions on the Chinese remainder theorem as a memory
saving trick. The work of GB was partly supported by  U.S.~DOE Grant
DE-FG02-90ER40542,  the research of MC was supported by  U.S.~DOE
Grant  DE-AC02-76CH00016, and the research of JL was partly supported
by the Swiss National Scientific Fund.  Some of the computations were
done on the Cray-YMP at the Supercomputing Computations Research
Institue at Florida State University and others used the Connection
Machine CM-2 at Thinking Machines Corporation in Cambridge, MA.

\vfill\eject
{\bf \noindent References}
\vskip .2in
\item{1.} K.~Binder, Physica 62 (1972) 508.

\item{2.} G.~Bhanot, J.~Stat.~Phys. 60 (1990) 55; G.~Bhanot and
S. Sastry, J. Stat. Phys. 60 (1990) 333.

\item{3.} M.~Creutz, Phys.~Rev.~B43 (1991) 10659.

\item{4.} {\it Phase Transitions--Cargese 1980,} edited by M.~Levy,
J.~Le Guillou, and J.~Zinn-Justin (Plenum, New York, 1982).

\item{5.} M.F.~Sykes, J.W.~Essam, and D.S.~Gaunt, J.~Math.~Phys. 6
(1965) 283; T.~de Neef and I.G.~Enting, J.~Phys. A10 (1977) 801;
I.G.~Enting, Aust.~J.~Phys.~31 (1978) 515; A.J.~Guttmann and
I.G.~Enting, Nucl.~Phys.~B (Proc.~Suppl.) 17 (1990) 328.

\item{6.}  G.S.~Pawley, R.H.~Swendsen, D.J.~Wallace and K.G.~Wilson,
Phys.~Rev.~B29 (1984) 4030; M.N.~Barber, R.B.~Pearson, D.~Toussaint
and J.L.~Richardson, Phys.~Rev.~B32 (1985) 1720; M.~Creutz, P.~Mitra,
and K.J.M.~Moriarty, J.~Stat.~Phys.~43 (1986) 823.

\item{7.} M.~Creutz, Phys.~Rev.~B45 (1992) 4650.

\vskip .5in
{\bf\noindent Figure Captions}
\vskip .2in
Fig. 1.  The ratio $r_E$ defined in Eq.~(6) in the vicinity of the
Ising critical point.  The series expansion for this quantity was Pade
approximated in $z=3u^2/(1+3u^2)$ as the ratio of two polynomials, and
the curves are labeled by the highest power of $z$ appearing in the
numerator.

Fig. 2.  The same as Fig.~1 but now for the ratio $r_\sigma$ in
Eq.~(8).

\vfill\eject
{\baselineskip=12pt
\vbox
{\noindent
Table I.  The low temperature expansion coefficients for the average
energy per unit volume.
\medskip
\settabs 2 \columns
\+ $i$  & $e_i$\cr
\smallskip
\+ 0 &  0 \cr
\+ 2 &  0 \cr
\+ 4 &  0 \cr
\+ 6 &  12 \cr
\+ 8 &  0  \cr
\+ 10 &  60 \cr
\+ 12 &  -84 \cr
\+ 14 &  420 \cr
\+ 16 &  -1,056 \cr
\+ 18 &  3,756 \cr
\+ 20 &  -11,220 \cr
\+ 22 &  37,356  \cr
\+ 24 &  -118,164 \cr
\+ 26 &  389,220 \cr
\+ 28 &  -1,261,932 \cr
\+ 30 &  4,163,592  \cr
\+ 32 &  -13,680,288 \cr
\+ 34 &  45,339,000  \cr
\+ 36 &  -150,244,860 \cr
\+ 38 &  500,333,916  \cr
\+ 40 &  -1,668,189,060 \cr
\+ 42 &  5,579,763,432  \cr
\+ 44 &  -18,692,075,820 \cr
\+ 46 &  62,762,602,860  \cr
\+ 48 &  -211,062,133,044 \cr
}
}

\vfill\eject
{\baselineskip=12pt
 \vbox
{\noindent
Table II.  Coefficients $c_{ij}$ for the expansion of the
magnetization. Here ${1\over 2}\langle 1-\sigma \rangle=\sum_{i,j}
c_{ij} u^{2i} \lambda^{j}$ where $u=e^{-2\beta}$ and
$\lambda=e^{-2\beta H}$.  Unlisted coefficients for $i \le 20$ all
vanish.
 \medskip
\settabs 6 \columns
\+ $i$  & $ j=1$ & 2 & 3 & 4 & 5 \cr

\+ 3 & 1 & 0 & 0 & 0 & 0 \cr
\+ 4 & 0 & 0 & 0 & 0 & 0 \cr
\+ 5 & 0 & 6 & 0 & 0 & 0 \cr
\+ 6 & 0 & -7 & 0 & 0 & 0 \cr
\+ 7 & 0 & 0 & 45 & 0 & 0\cr
\+ 8 & 0 & 0 & -108 & 12 & 0\cr
\+ 9 & 0 & 0 & 64 & 332 & 0\cr
\+ 10 & 0 & 0 & 0 & -1,314 & 240\cr
}
}

\bigskip
{\baselineskip=12pt
 \vbox
{
\settabs 6 \columns
\+ $i$ & $j=4$ & 5 & 6 & 7 & 8 \cr
\+ 11 & 1,620 & 2,130 & 108 & 0 & 0\cr
\+ 12 & -651 & -14,020 & 2,976 & 56 & 8\cr
\+ 13 & 0 & 27,660 & 9,450 & 2,646 & 0\cr
\+ 14 & 0 & -23,040 & -132,867 & 27,216 & 2,448\cr
\+ 15 & 0 & 7,031 & 387,444 & -9,520 & 36,976\cr
\+ 16 & 0 & 0 & -508,428 & -1,101,660 & 179,172\cr
\+ 17 & 0 & 0 & 320,220 & 4,722,564 & -848,904 \cr
\+ 18 & 0 & 0 & -78,904 & -8,833,328 & -7,580,660\cr
\+ 19 & 0 & 0 & 0 & 8,680,245 & 51,142,152\cr
\+ 20 & 0 & 0 & 0 & -4,397,652 & -130,897,242\cr
}
}

\bigskip
{\baselineskip=12pt
 \vbox
{
\settabs 5 \columns
\+ $i$ & $j=9$ & 10 & 11 & 12 \cr
\+ 14 & 216 & 0 & 0 & 0 \cr
\+ 15 & 1,143 & 240 & 0 & 0 \cr
\+ 16 &  49,896 & 3,960 & 264 & 36 \cr
\+ 17 & 360,450 & 41,310 & 7,260 & 0 \cr
\+ 18 & 547,236 & 672,670 & 73,216 & 12,960 \cr
\+ 19 & -12,320,586 & 2,368,080 & 773,025 & 138,744 \cr
\+ 20 & -35,804,700 & -6,147,840 & 6,632,208 & 1,220,220 \cr
}
}

\bigskip
{\baselineskip=12pt
 \vbox
{
\settabs 5 \columns
\+ $i$ & $j=13$ & 14 & 15 & 16\cr
\+ 18 & 1,248 & 0 & 0 & 0 \cr
\+ 19 & 9,516 & 1,596 & 0 & 0 \cr
\+ 20 & 311,688 & 32,760 & 2,520 & 240\cr
}
}

\bye